\documentclass{aa}

\usepackage{times}
\usepackage{graphics}
\usepackage{aabib99}

\def\ie{{\rm i.e.}}

\def\cc{\ifmmode{\,{\rm cm}^{-3}}\else{$\,{\rm cm}^{-3}$}\fi}
\def\cq{\ifmmode{\,{\rm cm}^{-2}}\else{$\,{\rm cm}^{-2}$}\fi}
\def\kms{\ifmmode{\,{\rm km}\,{\rm s}^{-1}}\else{km~s$^{-1}$}\fi} 

\def\HH{\ifmmode{\rm H_2}\else{$\rm H_2$}\fi}
\def\twCO{\ifmmode{\rm^{12}CO}\else{$\rm^{12}CO$}\fi} 
\def\thCO{\ifmmode{\rm^{13}CO}\else{$\rm^{13}CO$}\fi} 
\def\Cp{\ifmmode{\rm C^+}\else{$\rm C^+$}\fi} 
\def\CHp{\ifmmode{\rm CH^+}\else{$\rm CH^+$}\fi}
\def\CHthp{\ifmmode{\rm CH_3^+}\else{$\rm CH_3^+$}\fi} 
\def\HCOp{\ifmmode{\rm HCO^+}\else{$\rm HCO^+$}\fi} 

\newcommand{\emm}[1]{\ensuremath{#1}}   
\newcommand{\emr}[1]{\emm{\mathrm{#1}}} 

\newcommand{\Ntot}{\emm{N_\emr{tot}}}
\newcommand{\Tac}{\emm{\tau_\emr{ac}}}
\newcommand{\Dvtot}{\emm{\Delta v_\emr{tot}}}
\newcommand{\Dvpdf}{\emm{\Delta v_\emr{PDF}}}
\newcommand{\Dvmp} {\emm{\Delta v_\emr{mp}}}
\newcommand{\Dvem}{\emm{\Delta v_\emr{em}}}
\newcommand{\Dvabs}{\emm{\Delta v_\emr{abs}}}
\newcommand{\fvol}{\emm{f_\emr{v}}}
\newcommand{\fsur}[1][\mbox{}]{\emm{f_\emr{s#1}}}

\begin{document}

\thesaurus{09(09.05.1; 09.11.1; 09.13.2; 09.19.1; 02.20.1)}

\title{The elusive structure of the diffuse molecular gas: shocks or
  vortices in compressible turbulence?}

\author{J. Pety and {\'E}. Falgarone}

\offprints{J. Pety} \mail{pety@lra.ens.fr}

\institute{Radioastronomie, CNRS, UMR 8540, {\'E}cole Normale
  Sup{\'e}rieure, 24 rue Lhomond, F-75005 Paris, France}

\date{Received xxxx / Accepted xxxx}

\titlerunning{Structure of diffuse molecular gas}

\authorrunning{Pety \& Falgarone}

\maketitle %

\begin{abstract}
  
  The cold diffuse interstellar medium must harbor pockets of hot gas to
  produce the large observed abundances of molecular species, the formation
  of which require much more energy than available in the bulk of its
  volume. These hot spots have so far escaped direct detection but
  observations and modeling severely constrain their phase-space structure
  \ie{} they must have a small volume filling factor (a few \%), surface
  filling factors larger than unity with large fluctuations about average
  and comparable velocity structure in ``pencil beams'' and ``large
  beams''.
  
  The dissipation of the non-thermal energy of supersonic turbulence occurs
  in bursts, either in shocks or in the regions of large shear at the
  boundary of coherent vortices. These two processes are susceptible to
  generate localized hot regions in the cold medium.  Yet, it is of
  interest to determine which of them, if any, dominates the dissipation of
  turbulence in the cold interstellar medium.
  
  In this paper, we analyze the spatial and kinematic properties of two
  subsets of hydrodynamical compressible turbulence: the regions of largest
  negative divergence and those of largest vorticity and confront them with
  the observational constraints. We find that these two subsets fulfill the
  constraints equally well. A similar analysis should be conducted in the
  future on simulations of MHD turbulence.
  

\end{abstract}

\section{Introduction}

For several decades now, molecules have been detected in the diffuse
component of the cold neutral medium and these observations raise several
intriguing questions.  Firstly, the large observed abundances of \CHp{}
\cite{crane95:hrsiCH,gredel97:iCHsOBa} and \HCOp{} and OH
\cite{lucas96:pbsgHCOatecs} in the cold diffuse medium imply that
activation barriers and endothermicities of several thousands Kelvin be
overcome.  The formation of \CHp{} proceeds through the endothermic
reaction of \Cp{} with \HH{} ($\Delta E/k$=4640 K) and that of OH via the
reaction of O with \HH{} which has an activation energy $\Delta E/k$ =2980
K. In the diffuse gas, \HCOp{} forms from \CHthp, a daughter molecule of
\CHp. None of the large observed abundances can be explained by standard
steady-state chemistry in cold diffuse gas.  Pockets of hot gas must
therefore exist in the cold diffuse medium.

Secondly, many other molecules like CO, CS, SO, CN, HCN, HNC, H$_2$S,
C$_2$H have now been detected in absorption in front of extragalactic
continuum sources in local or redshifted gas (Lucas \& Liszt 1993, 1994,
1997; Liszt \& Lucas 1994, 1995, 1996; Wiklind \& Combes 1997, 1998).
\nocite{lucas93:pbommwmaBLL,lucas94:pbommwma,lucas97:mwodc}
\nocite{liszt94:mHCOedctdo,liszt95:mwmeaca,liszt95:mwmeaca}
\nocite{wiklind97:malhr,wiklind98:cmals} The lines of sight sample the
edges of molecular clouds or the diffuse medium, which corresponds to gas
poorly shielded from the ambient UV radiation field.  This derives from the
low excitation temperatures (often close to the temperature of the cosmic
background) measured in these transitions.  Yet, the molecular abundances
derived are close to those of dark clouds.

Thirdly, the spatial and velocity distribution of these molecules is highly
elusive. Observations of CO emission and absorption lines toward
extragalactic sources by Liszt \& Lucas \cite*{liszt98:COaecercs} show
that: {\it (i)} absorption and emission line profiles have comparable
linewidths whereas the projected size of the sampled volumes of gas differ
by more than four orders of magnitude (60 vs. 10$^{-3}$ arcsec), {\it (ii)}
the excitation temperature of the CO molecules changes only weakly across
the profiles and {\it (iii)} there are very few lines of sight with no
absorption line detected. The phase-space distribution of the CO--rich gas
in diffuse clouds must therefore have a surface filling factor close to
unity, and its velocity field must be as adequately sampled by a pencil
beam than by a large beam. As mentioned by the authors, these observations
suggest a one-dimensional structure for the molecular component, which
contrasts with the profuse small scale structure observed in emission.

Two mechanisms, operating at very different size scales, can trigger a {\it
  hot chemistry} in the cold diffuse medium and produce the observed
molecular abundances: magneto-hydrodynamical (MHD)
shocks~\cite{elitzur78,draine86:mhdsdc:ps,draine86:mhdsdc:lstro,pineau86:tsisms:fdc,flower98:ctsim:pal}
and intense coherent vortices, responsible for a non-negligible fraction of
the viscous dissipation of supersonic turbulence
\cite{falgarone95:iticigktdhpfgm,falgarone95:csitldic,joulain98:necdsit}.
This is so because the dissipation of the non-thermal energy of supersonic
turbulence is concentrated in shocks and in the regions of large shear at
the boundary of coherent vortices. In both cases, only a few \% of hot gas
on any line of sight across the cold medium is sufficient to reproduce the
observed column densities of molecules. This small fraction of hot gas
corresponds to about 6 MHD shocks of 10 \kms{} or 1000 vortices with
rotation velocity of 3.5~\kms{} per magnitude of gas (or $\Ntot=1.8\times
10^{21}$ \cq) of density $n \approx 30$ \cc{}
\cite{verstraete99:hcdm:ssHHrl}.

The main problem met with the models of individual C shocks or vortices is
that the predicted shift between neutral and ionized species is larger than
observed.  In this paper, we simply address the issue of the impact of the
line of sight averaging upon the resulting line profiles in a turbulent
velocity field. We investigate the spatial and velocity distribution of the
regions of high vorticity or high negative divergence (shocks) in a
simulation of compressible turbulence to test whether or not the
space-velocity characteristics of any of these subsets fulfill the
observational requirements.  We carry our analysis on the numerical
simulations of compressible turbulence of Porter, Pouquet \& Woodward
\cite*{porter94:ksdtdsf}. They are $512^3$ simulations of midly
compressible turbulence (initial rms Mach number=1). The time step analysed
here is $t=1.2 \Tac$ where \Tac{} is the acoustic crossing time. At that
epoch in the simulation, many shocks have survived producing density
contrasts as large as 40, but the bulk of the energy at small scales is
contained in the vortical modes.

These simulations are hydrodynamical and do not include magnetic field.
The impact of magnetic field upon the statistical properties of turbulence
may be less important than foreseen.  Recent simulations of MHD
compressible turbulence \cite{ostriker99:ksesgmc:ddt} show that the
dissipation timescale of MHD turbulence is closer than previously thought
to that of hydrodynamical turbulence.  Descriptions of the energy cascade
in models of MHD turbulence also predict that vortices must play an
important role~\cite{goldreich95:tist.II:sat,lazarian99:rwsf}. Magnetic
field does not prevent either the intermittency of the velocity field to
develop
\cite{brandenburg96,galtier98:imhdf,politano98:khemhdctolscf,politano98:dlstmf,politano95:mimhdt}.
In MHD as in hydrodynamic turbulence, shocks interact and generate vortex
layers, which are Kelvin-Helmholtz unstable and eventually form vorticity
filaments.
  
Another limitation of our study lies in the fact that the Reynolds number
of the simulations is small compared to that of interstellar turbulence.
Recent studies of high Reynolds number turbulence in Helium bring the
unexpected result that the statistical properties of the velocity field
have little dependence with the Reynolds number, at large Reynolds
numbers~\cite{tabeling96:pdfsflrnt}. We therefore believe that the analysis
presented here on hydrodynamical simulations has some relevance to the
understanding of interstellar turbulence.

We discuss the velocity structure of the two subsets (regions of high
vorticity or of high negative divergence) and compare in each case the
velocity samplings provided by pencil beams versus large beams
(Sect.~\ref{sec:vssivsct}). In Sect.~\ref{sec:sdhvds}, we discuss the
spatial distributions of the two subsets. In Sect.~\ref{sec:co}, we compare
the results derived from the numerical simulations with the observations.

\section{Velocity structure of subsets of intense vortices and shocks in
  compressible turbulence.}
\label{sec:vssivsct}

\subsection{Integrated profiles}

\begin{figure}
  \resizebox{\hsize}{!}{%
    \includegraphics*[1.6cm,19.5cm][10.3cm,26.5cm]{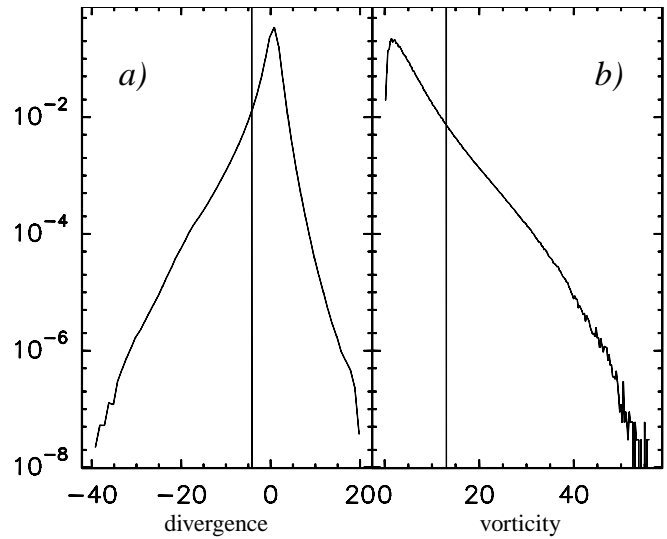}}
  \caption{PDFs of the values of {\it (a)} the positive and negative
    divergence and {\it (b)} the vorticity in the whole cube. The
    integrated numbers of points below (case {\it a}) above (case {\it b})
    each threshold indicated by a vertical bar are 3\%.}
  \label{fig:pdfs:ngv}
\end{figure}

We first compare the velocity distributions obtained with all the data
points in the simulation with those obtained by selecting only 3\% (or $4
\times 10^6$) of these data points. These subsamples are of three kinds: a
subset of randomly selected positions in space and the upper tails of the
distributions of the negative divergence and vorticity (see
Figs.~\ref{fig:pdfs:ngv}a and~\ref{fig:pdfs:ngv}b). In the following, for
the sake of simplicity, we will call them shocks and vortices.

The choice of 3\% of the points is a compromise between the value of
$\approx$ 1\% dictated by the chemical models and their confrontation to
the observations and a number large enough so that the statistical analysis
described below be meaningful. The actual values of the thresholds in the
vorticity or in the divergence correspond to this choice.  As said above,
many of the molecular species observed in the cold diffuse medium, require
large temperatures for their formation. Below gas temperatures of the order
of 10$^3$ K, there is an exponential cut-off of the speed or the
probability of these chemical reactions. Since the heating sources are
either the viscous dissipation enhanced in the layers of large velocity
shear at the boundary of coherent vortices \cite{joulain98:necdsit}, or the
ion-neutral streaming in the magnetic precursor of the shocks
\cite{flower95:ntsigmhds}, the thresholds in vorticity and in negative
divergence correspond to temperature thresholds. It is only above such
thresholds in negative divergence (shock velocity) or vorticity (shear),
that the {\it hot chemistry} required by the observations can be triggered
in the cold diffuse medium. As seen on Fig.~\ref{fig:pdfs:ngv}, the
selected thresholds for the vorticity and the divergence fall in the
non-Maxwellian (non-Gaussian) tails of each distribution. The sets of
structures that we describe therefore belong to the non-Maxwellian
(non-Gaussian) tails of the corresponding distributions.
 
\begin{figure}
  \resizebox{\hsize}{!}{%
    \includegraphics*[1.5cm,20.0cm][10.3cm,26.65cm]{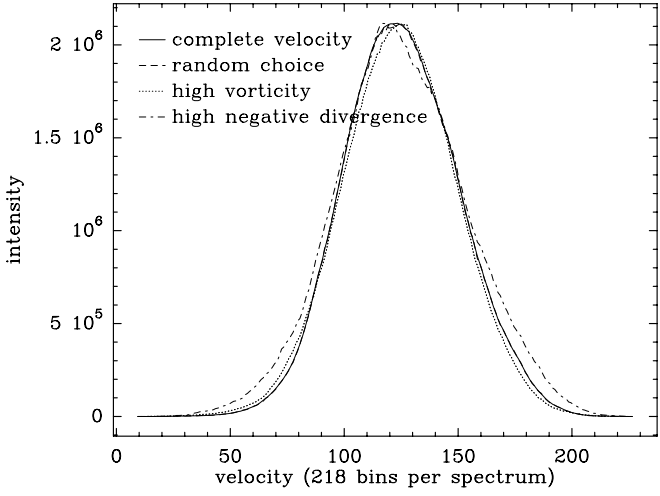}}
  \caption{Velocity distributions obtained with the full velocity field and
    all the points in the simulation (solid), with 3\% of the data points
    randomly selected in space (dashed), with the 3\% of the points which
    have the largest vorticity (dotted) and with the 3\% of the points
    which have the largest negative divergence (dot-dashed). The intensity
    scale corresponds to the number of points in each of the 218 velocity
    bins. The velocity scale is the bin number.}
  \label{fig:vd}
\end{figure}

The velocity distributions of each subset (normalized to their peak value)
is shown in Fig.~\ref{fig:vd} together with that of the whole cube.  They
are remarkably similar. The only subset which provides a slightly (10\%)
broader spectrum is that built on the shocks.  A tracer passively advected
in a turbulent flow (subset of spatially random positions) would therefore
carry the characteristics of the turbulent field of the bulk of the volume,
even though its volume filling factor is as small as a few \%.

\subsection{Velocity widths of pencil beam and large beam line profiles.} 

\begin{figure}
  \resizebox{\hsize}{!}{%
    \includegraphics*[5.0cm,4.1cm][15.1cm,21.9cm]{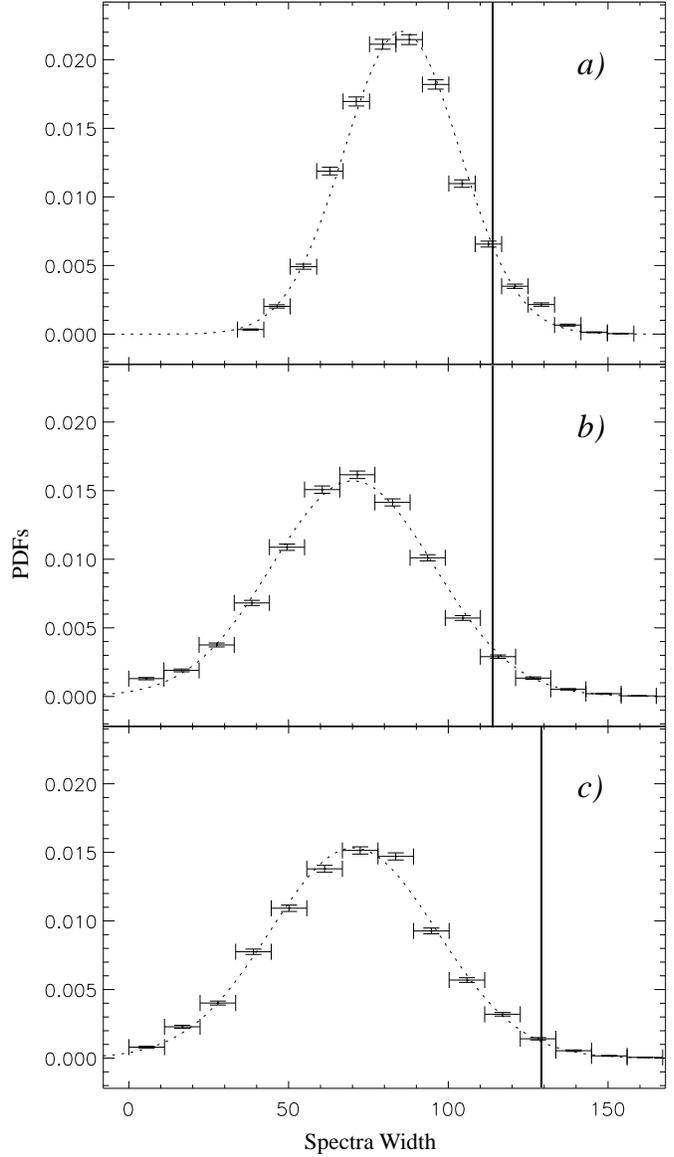}}
  \caption{PDFs of the velocity widths of the $1.6 \times 10^4$ individual
    spectra, each sampling $4 \times 4 \times 512$ pixels, and therefore
    mimicking spectra obtained with a beam 128 times narrower than that
    obtained by integrating the emission of the whole cube. The spectra are
    built on {\it (a)} the complete velocity field, and on only 3\% of the
    data points selected {\it (b)} as the most intense vortices and {\it
      (c)} as the most intense shocks. The vertical bar, in each panel
    indicates the width, at the same level, of the integrated spectrum over
    the whole cube.}
  \label{fig:pdfs:sw}
\end{figure}

\newcommand{\e}[1]{\ensuremath{\times 10^{#1}}}
\newcommand{\DivV}{\ensuremath{|\nabla .v|}}
\begin{table*}
  \begin{flushleft}
  \caption{Comparison of integrated and individual spectra.}
  \label{tab:ciis}
    \begin{tabular}{lrcccccc}
      \hline
      \noalign{\smallskip}
      & &
      \multicolumn{2}{c}{Integrated spectrum} &
      \multicolumn{4}{c}{Individual spectra} \\
      \cline{5-8}
      Sample & level &
      \fvol{} & \Dvtot{} & $\fvol'$ & \Dvmp{} 
      & ${\Dvtot - \Dvmp \over \Dvtot }$ & 
      \Dvpdf{} \\
      \noalign{\smallskip}
      \hline\noalign{\smallskip}
      full field     &  8\% &       1 & 114 & 6\e{-5} & 85 & 0.25 & 43 \\
                     & 16\% &         &  97 &         & 73 & 0.25 & 43 \\
      \\
      large $\omega$ &  8\% & 3\e{-2} & 113 & 2\e{-6} & 70 & 0.38 & 61 \\
                     & 16\% &         &  96 &         & 64 & 0.33 & 62 \\
      \\
      large \DivV{}  &  8\% & 3\e{-2} & 130 & 2\e{-6} & 70 & 0.46 & 61 \\
                     & 16\% &         & 110 &         & 62 & 0.44 & 62 \\
      \noalign{\smallskip}
      \hline
    \end{tabular}
  \end{flushleft}
\end{table*}

We now compare the velocity width of synthetic spectra obtained with a
large beam (\ie{} profiles observed in emission) to those obtained in a
pencil beam (\ie{} profiles observed in absorption). To do so we have
computed the velocity distribution in $4\times 4\times 512$ subsamples of
the whole cube and of the two subsets of largest vorticity and largest
divergence. Under the approximation that the line radiation is optically
thin \cite{falgarone94:sstc}, we consider that these velocity distributions
capture the main characteristics of the line profiles. The velocity
distributions will therefore be called profiles (or spectra) for simplicity
in what follows. A total number of $1.6 \times 10^4$ individual spectra are
therefore obtained across the face of the cube, for each set.  To achieve
the comparison of these individual spectra (surrogates of pencil beam
spectra) with the total spectrum (surrogate of the large beam spectrum), we
use a width computed at levels 1\%, 8\%, 16\% and 32\% of the local peak of
the velocity distribution (spectrum). Radiative transfer affects more
severely the velocity domains where the crowding is the largest (the peaks
of the velocity distributions) and this is the reason why we analyze the
velocity coverage of each distribution at levels low enough for radiation
to be reasonably assumed optically thin.

We have computed the probability distribution functions (PDF) of the widths
of the individual profiles at the four levels quoted above.  They are shown
on Fig.~\ref{fig:pdfs:sw} at the 8\% level for the whole velocity field and
the large vorticity and divergence subsets.  The dotted curve shows the
Gaussian distributions with the same mean and dispersion. On each panel,
the vertical bar indicates the width of the large beam profile at the same
level (\Dvtot{} in Table~\ref{tab:ciis}). The width (\Dvpdf{}) and peak
(\ie{} most probable value, \Dvmp{}) of the PDFs are given in
Table~\ref{tab:ciis} for the 8\% and 16\% levels and differ by less than
20\% with the level selected. The volume filling factors \fvol{} and
$\fvol'$ represent the fraction of points included in each sample.

In all cases, the pencil beam spectra are narrower than the large beam
spectrum but by only 25 to 45\% (column 7 of Table~\ref{tab:ciis}) while
the projected sizes of the sampled volumes differ by a factor 128.  This is
due to the fact that although the projected size of the volumes sampled by
the pencil beams are small, the depth along the line of sight has remained
unchanged (512). Along one dimension at least, the pencil beam samples the
velocity over large scales.  We have made similar computations for $8\times
8\times 512$ and $32\times 32\times 512$ pencil beams and the widths of the
corresponding velocity distributions remain almost the same. For this
reason we believe that the results for a pencil beam as small as those of
actual observations compared to the large beam (\ie{} 60"/0.001"=6$\times
10^4$, a ratio which cannot be achieved by any direct numerical simulation)
would not be significatively different. As long as the depth of the medium
sampled by the line of sight is large, the velocity signature of the pencil
beam profile remains close to that of these large scales.

Then it is interesting to note that the differences between the subsets of
strong shocks and intense vortices are not marked. On average, vortices
produce pencil beam profiles closer to the large beam profile than do the
shocks, but the effect in the present simulation is small.

\subsection{PDFs of line centroid increments}

\begin{figure*}
  \resizebox{12.0cm}{!}{%
    \includegraphics*[3.0cm,3.0cm][15.1cm,26.7cm]{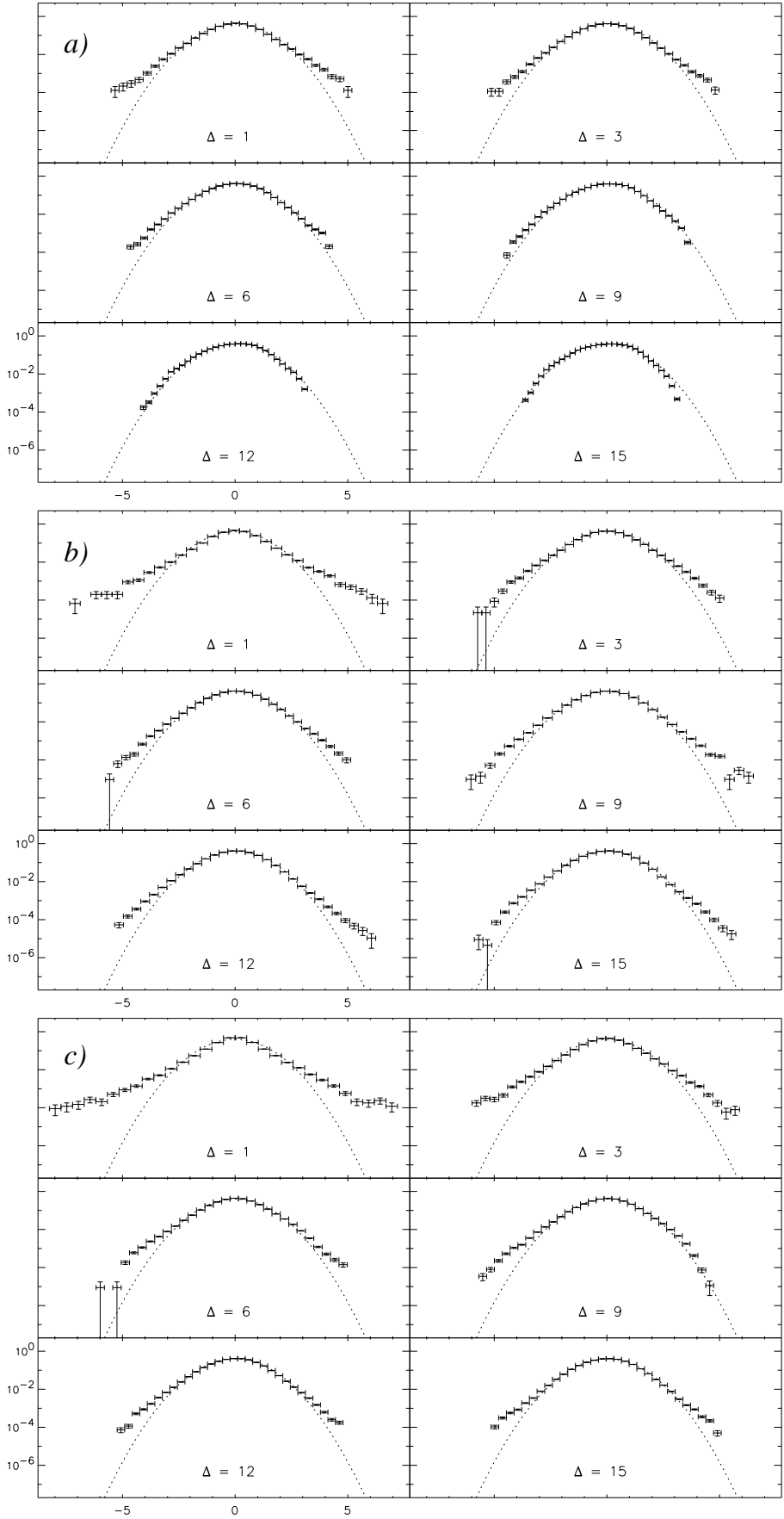}} \hfill
  \parbox[b]{55mm}{%
    \caption{PDFs of the line centroid increments for different lags
      $\Delta$ and {\it (a)} the full velocity field, {\it (b)} the subset
      of most intense vortices and {\it (c)} the subset of positions of
      largest negative divergence.}}
  \label{fig:pdfs:ci}
\end{figure*}

We have computed the line centroid of each spectrum across the face of the
cube according to the method described in Lis et
al.~\cite*{lis96:splcvcvict}. Fig.~\ref{fig:pdfs:ci} displays the
probability distribution functions of the transverse increments of these
centroids for several lags.  These PDFs are normalized to the rms
dispersion of the increments.  The comparison of the PDFs obtained for the
full velocity field (Fig.~\ref{fig:pdfs:ci}a) with those obtained with the
3\% most intense vortices (Fig.~\ref{fig:pdfs:ci}b) and shocks
(Fig.~\ref{fig:pdfs:ci}c) shows that, at the smallest lag, the non-Gaussian
tails of the PDFs extend much further for the latter subsets that for the
former.  Non-Gaussian tails disappear at $\Delta=9$ for the full velocity
field while they persist up to $\Delta=15$ for the regions of large
divergence and large vorticity. The non-Gaussian tails of the PDFs of line
centroid increments have been shown to be associated with regions of
enhanced vorticity in turbulence~\cite{lis96:splcvcvict}. This result shows
that both the subsets of shocks and vortices exhibit pronounced
non-Gaussian behaviour in those PDFs.
 
\begin{figure*}
  \resizebox{\hsize}{!}{%
    \includegraphics*[1.0cm,3.0cm][20.0cm,26.5cm]{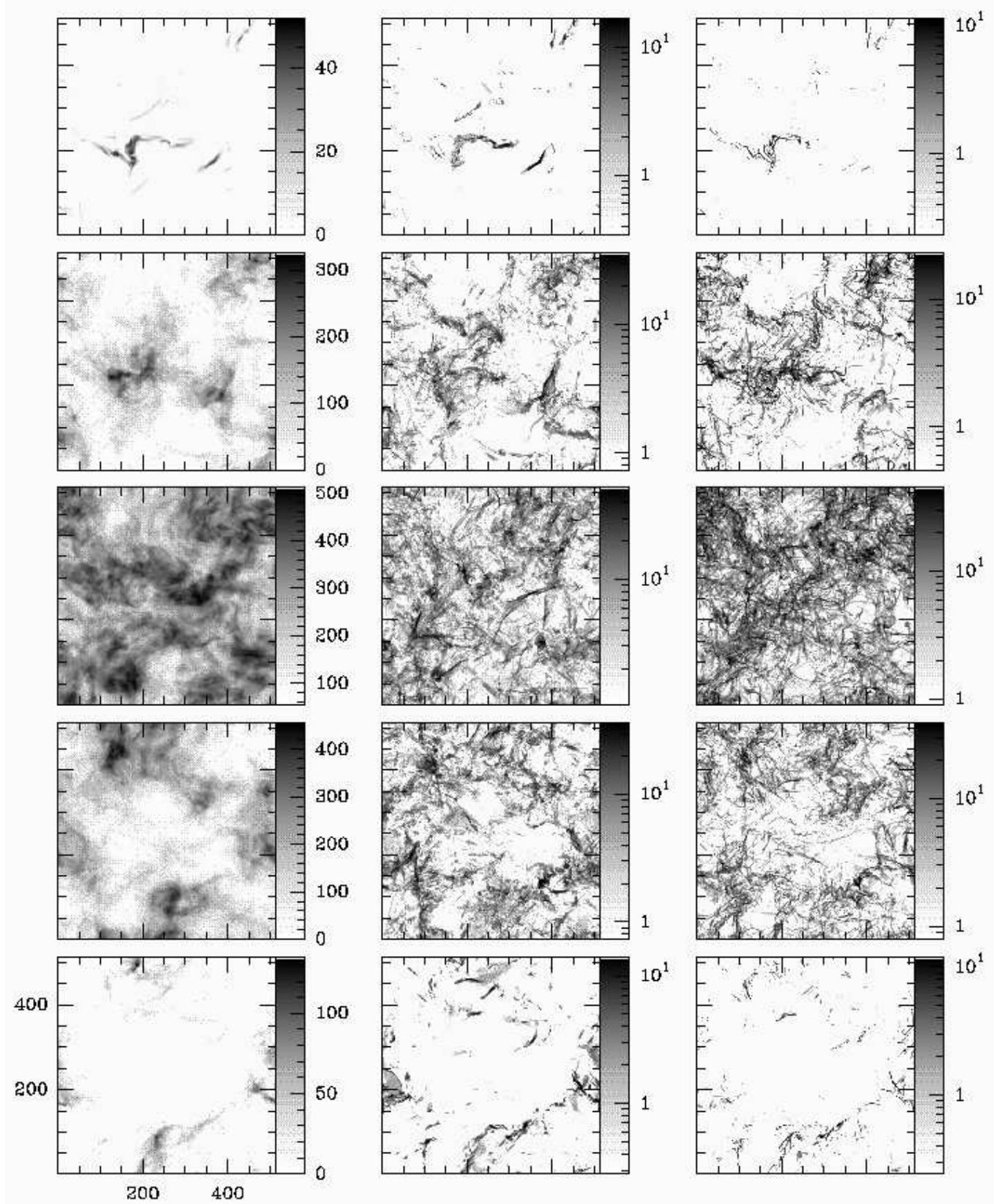}}
  \caption{Channel maps of the surface filling factor on each line of sight
    in {\it (a)} the complete velocity field, {\it (b)} the subset of the
    most intense vortices and {\it (c)} the subset of the largest negative
    divergence. The color scales are different for each velocity range and
    do not cover the whole dynamic range for the vortices and the shocks.
    From top to bottom panel, the values of the actual maxima are (52, 326,
    512, 457, 135) for the complete velocity field, (23, 46, 90, 79, 24)
    for the vortices and (36, 83, 116, 73, 30) for the shocks.}
  \label{fig:cmsff}
\end{figure*}

\begin{figure*}
  \resizebox{\hsize}{!}{%
    \includegraphics*[1.5cm,7.5cm][19.5cm,22.5cm]{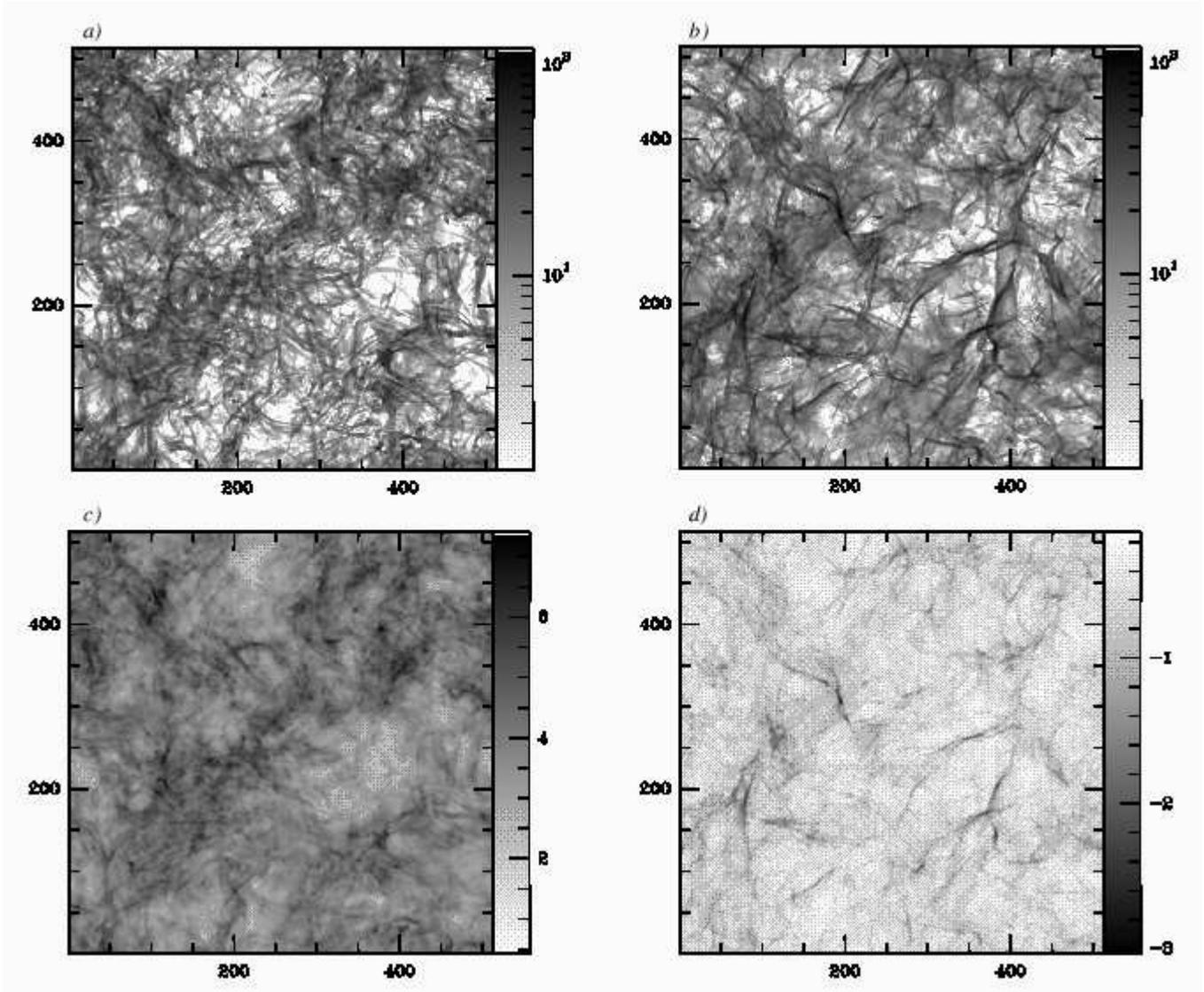}}
  \caption{Maps of the surface filling factors of {\it (a)} the 
    most intense vortices {\it (b)} the regions of largest negative
    divergence. Maps of line--of--sight average of {\it (c)} the vorticity
    modulus and {\it (d)} the negative divergence.}
  \label{fig:msff}
\end{figure*}

\section{Spatial distribution of the high vorticity/divergence
  subsets}
\label{sec:sdhvds}

\subsection{Channel maps}

The three channel maps of Fig.~\ref{fig:cmsff} display the surface filling
factor of the data points in five adjacent velocity intervals for the whole
cube and the two subsets of vortices and shocks. To optimize the visibility
of the weakly populated regions, the color scale does not span the whole
dynamic range and is not the same in each velocity interval. At almost all
velocities, the surface filling factors reach larger values for the shocks
than for the vortices (see caption of Fig.~\ref{fig:cmsff}). Thus the
shocks are more clustered in space than the vortices. Shocks also appear as
thicker (or more extended) projected structures than vortices. Both subsets
have a much more pronounced filamentary texture than the whole velocity
field.  Many small scale filaments are visible with lengths reaching a
significant fraction of the integral scale.  This result has been already
recognized for the regions of large vorticity by Vincent \& Meneguzzi
\cite*{vincent91,vincent94} or She, Jackson \& Orszag \cite*{she90} in
incompressible turbulence and by Porter, Woodward \& Pouquet
\cite*{porter98:irsdctf} in compressible turbulence.

For the bulk of the structures, there is no coincidence between the
patterns delineated by the shocks and the vortices. There are exceptions,
though, like the structures seen in the extreme velocity channels. The
small scale patterns delineated in these channels by the regions of large
vorticity or large divergence closely follow each other, although the
details do not exactly coincide.  It is remarkable that all the small scale
patterns seen in the extreme velocity channels in the maps of large
vorticity and divergence are also those of the full velocity field
(Fig.~\ref{fig:cmsff}a): in the first channel (top left), the data points
of the full velocity field are either shocks or vortices or both.

These maps also show that shocks and vortices are not randomly distributed
but are clustered in space and in velocity. There are large voids with only
a few vortex or shocks on the line of sight. The contrasts are large as
indicated by the peak values given in the caption of Fig.~\ref{fig:cmsff}.

\subsection{Integrated maps and surface filling factors}

We now turn to the spatial distribution across the face of the cube of the
two subsets of shocks and vortices. The set of randomly selected positions
does not have any significant spatial structure. It is therefore not shown.
The spatial distribution of the surface filling factor \fsur{} is shown in
Fig.~\ref{fig:msff} for the vortices and the shocks.

Fig.~\ref{fig:msff} shows that the contrasts are large ($\approx 100:1$) in
the two subsets for which the average surface filling factor per pixel is
only 15.3 (\ie{} 3\% of the data points over 512$^2$ pixels). There are
only 5\% and 2\% empty lines of sight for the vortices and the shocks
respectively, \ie{} the surface filling factors of the regions of large
vorticity or large divergence are almost everywhere larger than unity,
although they fill only 3\% of the whole volume. These numbers depend on
the thresholds selected for the vorticity and divergence.

The maps of the filling factors (Figs.~\ref{fig:msff}a and~\ref{fig:msff}b)
are very similar to the maps of the vorticity and of the negative
divergence (Figs.~\ref{fig:msff}c and~\ref{fig:msff}d) computed with the
whole data cube. In particular the maxima in vorticity and in negative
divergence are those of the surface filling factors of the upper tails of
the corresponding distribution. It means than for the vortices and for the
shocks, the origin of the maxima observed in projection is the crowding of
rare events populating the upper tails of the distributions.

\begin{figure}
  \resizebox{\hsize}{!}{%
    \includegraphics*[1.5cm,20.0cm][10.3cm,26.6cm]{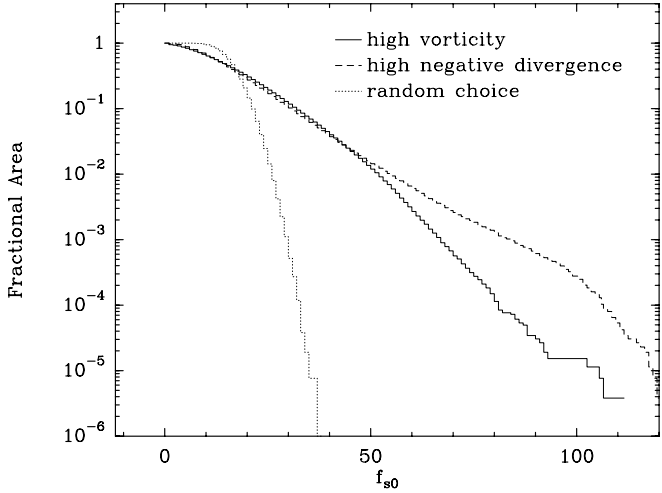}}
  \caption{Fractional area covered by the structures with a surface
    filling factor above a threshold \fsur[0] in the high vorticity (solid)
    and high negative divergence (dashed) subsets. The same curve for the
    randomly selected positions is the dotted curve.}
  \label{fig:fa}
\end{figure}

We have computed the fractional area covered by structures with surface
filling factors larger than a threshold \fsur[0]. The dependence of this
fractional area on \fsur[0] is shown in Fig.~\ref{fig:fa}. Unlike randomly
selected positions (dotted curve), the high vorticity/divergence
distributions exhibit structures with large filling factors which cover
only a tiny fraction of the total area.  The large crowdings of shocks are
more numerous than those of vortices, but the bulk of the ensemble of
vortices (those which fill most of the surface, $\fsur[0]<50$) cover a
fractional area slightly larger that the bulk of the shocks.

\section{Comparison with observations}
\label{sec:co}

Statistics on molecular absorption lines in the cold diffuse medium are
still sparse. We have used the data set of Liszt \& Lucas
\cite*{liszt98:COaecercs} to compare the velocity widths of 9 pairs of
\thCO(1-0) and \thCO(2-1) absorption and emission lines toward
extragalactic sources. In 5 cases (B1730, B2013, B2200 and B2251), the
\twCO{} lines are simple enough that a comparison between the emission and
absorption profiles is meaningful. Out of the 14 pairs, two exhibit
emission and absorption profiles of equal widths, three have absorption
profiles larger than the emission profiles (this is a possible
configuration according to the simulations, see Fig.~\ref{fig:pdfs:sw}) and
nine have absorption narrower than emission profiles. The differences are
not large and the average value of the 9 relative width differences
($\Dvem-\Dvabs/\Dvem$) is only 0.34. This number is very close to the
values listed in Column 7 of Table~\ref{tab:ciis} for the subset of large
vortices (although the difference with that of large divergence is not
significant).

Another interesting property of the molecular line absorption measurements
is the large scatter of column densities detected for a given amount of gas
sampled by the line of sight, \Ntot{}. On average, the observed column
density of \CHp{} for instance increases almost linearly with \Ntot{} (see
references in Joulain et al. \cite*{joulain98:necdsit}), but the scatter of
the values at a given value of \Ntot{} is large. Scatters of about a factor
10 are found in merging the samples of Crane et al.  \cite*{crane95:hrsiCH}
and Gredel \cite*{gredel97:iCHsOBa}. It means that the spatial distribution
of the structures bearing these molecules is highly heterogeneous, although
{\it on average}, the larger the total column density of gas sampled, the
larger the column density of molecules.  This characteristic may be brought
together with the fact that the surface filling factor of the most intense
vortices and shocks have large fluctuations (a factor $\sim$ 10) about
their average value (15) (Figs.~\ref{fig:msff}a and~\ref{fig:msff}b).

\section{Conclusion}

The regions of largest vorticity or largest divergence in compressible
turbulence are small subsets of a whole turbulent velocity field but we
have shown that, despite their small volume filling factor (here
$\fvol=0.03$):

(i) they sample the whole velocity field,

(ii) pencil beam samplings across these subsets have velocity coverages
almost as broad as those obtained with large beams. The values (35\% to
45\%) and the signs of these differences are consistent with the
observations.

(iii) the PDFs of centroid velocity increments built on these subsets have
more extended non-Gaussian wings than those of the full velocity field.
Their intermittent characteristics subsist at larger lags than for the full
velocity field.

(iv) copious small scale structure with large contrasts is seen in the maps
of the surface filling factor of the regions of large vorticity and large
divergence.  These contrasts are of the same order of magnitude as those
observed in the column densities of \CHp{} for instance for a given column
density of sampled gas.

(v) the surface filling factor of the subsets of high vorticity/divergence
are almost everywhere larger than unity even though their volume filling
factor is as small as 3\% in the subsets studied here. Vortices are
slightly more efficient than shocks at covering the sky: they tend to be
more numerous, and to form smaller structures which are less clustered in
space.

In summary our study confirms that mild shocks as well as intense vortices
could be the subsets of the cold diffuse medium enriched in molecules and
responsible for the molecular absorption lines detected in the direction of
extragalactic sources.

Numerical simulations of hydrodynamical turbulence are not ideally suited
to test the properties of such subsets, but one-fluid simulations of MHD
turbulence are not ideal either because of the importance of the
ion-neutral streaming in the triggering of hot chemistry, whether in MHD
shocks or in magnetized vortices. More detailed predictions of the
phase-space structures of the subsets of turbulence where hot chemistry is
activated in the cold diffuse medium require calculations of MHD
compressible turbulence which take ion-neutral drift into account.

\begin{acknowledgements}
  We thank D. H. Porter, A. Pouquet, and P. R. Woodward for providing us
  with the result of their hydrodynamic simulation and our referee, A.
  Lazarian, for his helpful comments.
\end{acknowledgements}

\bibliography{ms9327}
\bibliographystyle{aabib99}

\end{document}